\input epsf.tex
\documentstyle[aps]{revtex}
\def \beq{\begin{equation}}
\def \eeq{\end{equation}}
\def \ps{\psi}
\def \pb{\bar \psi}
\def \gi{\gamma_{i}}

\def \g5{\gamma_{5}}
\begin{document}
\baselineskip=24pt
\vspace{1.5cm}
\begin{center}
\bf{ Axial Anomaly and Ginsparg-Wilson fermions in the Lattice Dirac Sea
picture}

\vspace{1cm}
\rm{Srinath Cheluvaraja} \\
\it{Dept. of Physics and Astronomy, Louisiana State University,
Baton Rouge, LA, 70808} \\

\rm{N.D. Hari Dass} \\
\it{Institute of Mathematical Sciences,
Chennai, 600113} \\
\end{center}

\noindent{\bf{ABSTRACT}}\\
The axial anomaly equation in 1+1 dimensional QED is obtained on the lattice
for fermions obeying the Ginsparg-Wilson relation. We make use of the 
properties of the Lattice
Dirac sea to investigate the connection between the anomaly and the 
Ginsparg-Wilson operator in the Hamiltonian picture. The correct anomaly is
reproduced for gauge fields whose characteristic time is much larger than
the lattice spacing, which is the regime where the adiabatic approximation
applies. A non-zero Wilson $r$ parameter is
necessary to get the correct anomaly. The anomaly is shown to be independent
of $r$ for $r>0.5$. The generalization to 3+1 dimensions is also discussed.

\vspace{0.5cm}
\begin{flushleft}
\end{flushleft}

\newpage

The lattice regularization is one of the few non-perturbative methods available
for defining quantum field theories. Lattice gauge theories have revealed
many interesting features of gauge theories that are not easily visible
in the usual perturbative approach. Nevertheless, the lattice regulator
has proved problematic if fermions have to be incorporated into the theory.
A naive discretization of the fermionic theory suffers from the replication
of fermion modes due to 
the "doublers". The doublers are degenerate in energy
with the originally introduced 
 fermions and though they have lattice momenta of the order of the cut-off ($1/a$
in lattice theories, $a$ is the lattice spacing), they mimic ordinary low energy
fermions. These doubler modes
cannot be ignored as they participate in physical processes, for instance
they can be pair created, and can affect the value of physical quantities--
such as the free energy. The first method to handle these doublers was given in
\cite{wils} and it uses an additional term in the action --the Wilson term--
to lift the degeneracy of the fermions, thereby 
decoupling the doublers in the continuum limit. However,
this method has the disadvantage of explicitly breaking chiral symmetry, and
hinders the study of dynamical questions related to chiral symmetry breaking.
A cure for the doubling problem that explicitly breaks chiral symmetry
also makes the lattice regularization
of chiral gauge theories, such as the standard
model, much more difficult. The Nielsen-Ninomiya no-go theorem \cite{nogo} decrees
that any chirally symmetric 
lattice Hamiltonian satisfying general properties like locality
and hermiticity must result in a replication of fermion species.
This theorem seems to suggest the impossibility of defining
undoubled fermions on the lattice without breaking chiral
symmetry.
Recently, however, alternative methods for tackling these problems have
emerged. One of them \cite{has}
uses the so called Ginsparg-Wilson relation \cite{gwils}
for Dirac fermions.The
Ginsparg-Wilson (G-W) operator is obtained by the application of block-spin
transformations to a chirally invariant
Dirac operator, and which therefore suffers from fermion doubling, using a 
{\bf chirally non-invariant} blocking kernel. Although the G-W operator is
not chirally invariant, it contains the information
of chiral symmetry because it has been obtained after blocking
a chirally invariant lattice action.
Its construction, by 
a renormalization group transformation of a chirally invariant action,
is bound to leave the low energy properties related to the 
chiral symmetry unchanged.
This approach of formulating lattice fermions has led to many recent
developments \cite{oth}, such as, lattice formulations of chiral symmetry,  the search
for chiral gauge theories on the lattice, methods of defining a
lattice  topological
charge, and formulation of lattice index theorems etc.

The G-W operator has to satisy the following relation
\beq
D\g5+\g5 D=a D\g5 D
\quad .
\eeq
(There are different versions of the G-W relation depending on the precise
form of the blocking kernel used. The above form is one of the simpler
ones and is sufficient for the ensuing discussion. Here $a$ is the lattice
spacing.
)
The G-W operator clearly does not satisfy chiral symmetry (because
$\{D,\g5\}\ne 0$). 
Even though the G-W operator seems to share the properties of a
chirally noninvariant mass term,
it is a milder way to break the chiral symmetry on the lattice. This is
because it is obtained by blocking a chirally symmetric action, although
using a chirally non-invariant kernel. The low energy properties of the
G-W operator on the lattice are the same as those of the chirally
symmetric action. 

Another approach to problems of chirality on the lattice
is the overlap approach introduced in \cite{neub}. 
It
captures many essential elements of domain wall fermions \cite{kap} as well as 
the one in \cite{fro} which requires an infinite number of auxiliary
fields. The chiral determinant is expressed as an overlap of the ground
states of two many body Hamiltonians and a construction of chiral gauge
theories involves regularizing the overlap \cite{chi}. 
Though the original Ginsparg-Wilson approach and the overlap approach 
appear to have 
nothing in common, the overlap operator (which appears in the
Hamiltonian)
has been shown to satisfy the
Ginsparg-Wilson relation \cite{ngw}. The overlap is also a real time approach since
it involves the quantum mechanical scalar product of the ground states of two
Hamiltonians. This approach has led to many further studies of chirality
on the lattice \cite{stud}.

Another useful way of looking at the fermion doubling problem on the lattice is to
look at the chiral anomaly structure of the lattice theory and to see what
it yields in the continuum limit. A symmetry is said to be anomalous if
it is no longer present in the quantum theory although it is present in the
classical theory. The anomaly manifests itself by a non-conservation of a
classically conserved charge.
Anomalies are an inescapable part of some quantum field theories and
have many important physical consequences. Their origin is related to
the problem of regularizing amplitudes in quantum field theories while maintaining
their invariances.
In the path integral formulation of quantum field theory they arise
because of the non-invariance of the measure of the path integral \cite{kaz}.
As is well known 1+1 dimensional QED has a chiral anomaly when massless
fermions are present. The anomaly arises because it is not possible to
find a regularization of the gauge theory which maintains both the gauge
invariance and  chiral symmetry. A simple way of demonstrating the anomaly
in 1+1 dimensions is by using a gauge invariant
point-split definition of the axial
vector current \cite{split} which can be seen not to be conserved.
There are no anomalies in the naive latticisation because it is a gauge 
invariant regulator which also maintains chiral symmetry,
but it is impossible to put only fields of one chirality on the lattice. 
This is consistent with the fact that no regularisation exists which
simultaneously preserves gauge invariance and chiral symmetry for arbitrary
matter content. The continuum limit of the lattice theory, 
on the other hand, must be able to reproduce the
correct anomaly structure of a given theory. The naive fermionic lattice action coupled to gauge fields
gives an anomaly free
theory in the continuum limit because the anomalies are  cancelled between the
naive and doubled modes \cite{smit}. 
It was shown in \cite{smit} that the Wilson term reproduces the
correct anomaly on the lattice provided the symmetry breaking parameter
$r\ne 0$, the anomaly in the continuum limit being
given by the co-efficient of the Wilson term. This
is quite a surprising result because the Wilson term explicitly breaks the
chiral symmetry but yet reproduces the anomaly which is essentially a
quantum mechanical breakdown of the classical chiral symmetry.

As stressed by Nielsen and Ninomiya , and Peskin, the Hamiltonian 
formulation provides a much clearer physical picture of
the anomaly in terms
of the energy level shifting of the
filled Dirac sea \cite{niels}( for a very clear exposition see also 
\cite{holger}). In this picture
the anomaly arises because
pairs of net chirality are pumped out of the infinitely filled Dirac sea.
If one tries to transcribe this picture on the lattice, as was done by
Ambjorn et al \cite{amb}, one finds
that the lattice Dirac sea is always finite and the anomaly always gets
cancelled by the doubler modes in the absence of the Wilson term\cite{smit,amb}.
The physical picture of the anomaly presented in \cite{niels} can be applied 
on the lattice with a Wilson mass term. The role of the Wilson mass term is to
suppress the contributions to the chiral charge coming from the doubler
modes resulting in a non-zero anomaly on the lattice
\cite{amb}.

Our aim is to carry out a similar analysis for fermions satisfying the G-W relation
and to see how a non-zero anomaly comes about on the lattice. 
This should complement
the derivation of the anomaly from the Ginsparg-Wilson action in the Euclideanised
formalism where the anomaly is showed to arise out of the measure \cite{luscher}.
We mention here that the axial anomaly in 1+1 dimensional QED is
also reproduced in the overlap formulation \cite{neub}. Our derivation, apart
from being quite different from the methods employed in \cite{luscher,neub},
also highlights the role played by the Wilson term in giving
the correct anomaly.
The discussion will be in the
Hamiltonian framework and we will derive the anomaly equation for
the abelian theory in the 1+1 dimensions. We will then comment on the extension
of this picture to 3+1 dimensions.

The Lagrangian density for a Dirac fermion in Minkowski space is given by
\beq
L(\psi, \bar \psi)=\bar \psi(i\gamma_{\mu}\partial_{\mu}-m)\psi
\quad ;
\eeq
the gamma matrices satisfy
$\gamma_{0}^{\dagger}=\gamma_{0},\ 
\gamma_{i}^{\dagger}=-\gamma_{i}$, and obey the relation
$\{\gamma_{\mu},\gamma_{\nu}\}=-2g_{\mu \nu}$.
The metric $g_{\mu \nu}$ is $diag(1,-1,-1,-1)$.
$\bar \psi(x)$ denotes the relativistic adjoint
$\psi^{\dagger}(x)\gamma_{0}$.
We shall use the Weyl representation for the gamma matrices.
The Hamiltonian density is given by
\beq
H=\bar \psi(x)(-i\gamma_{i}\partial_{i}+m)\psi(x) \quad
\eeq
Before we discuss the GW fermions in the Hamiltonian picture, it is instructive
to briefly review how the unwanted doublers are handled in the Euclidean
formalism with the help of the Wilson mass term.In Euclidean space 
the Lagrangian density becomes
\beq
L(\psi, \bar \psi)=\bar \psi(-i\gamma_{\mu}\partial_{\mu}-m)\psi
;
\eeq
the Euclidean gamma matrices satisfy
$\{\gamma_{\mu},\gamma_{\nu}\}=-2\delta_{\mu \nu} \quad $, and
$\gamma_{\mu}^{\dagger}=
-\gamma_{\mu}$.
The naive lattice discretization of the Dirac action for a
massless fermion 
in Euclidean space is given by
\beq
S=\sum_{x\ i}\frac{-i}{2a} \pb(x) \gi(\ps(x+i)-\ps(x-i))
\quad .
\eeq
In momentum space (in $d$ dimensions) this becomes
\beq
S=\int_{BZ} d^{d}k \  \bar \psi(k)(\sum_{i=1}^{d}(\frac{\sin(k_{i}a)}{a})\gi)
\psi(k)
\quad .
\eeq
$BZ$ denotes the range of integration to be the $d$ dimensional Brillouin
zone of the lattice.
As is well known, the above discretization suffers from the presence of
additional fermions at the corners of the Brillouin zone
($2^{d}$ in $d$ Euclidean dimensions)
leading to $2^{d}$
fermions in the $a\rightarrow 0$ limit.
A method for eliminating the unwanted fermions
is to give them
very high masses in the continuum limit. The oldest way of
achieving this is by adding a Wilson term to the massless action
\beq
S_{w}= -\frac{r}{2a}\sum_{x i}\pb(x)(\ps(x+i)+\ps(x-i)-2\ps(x))
\quad .
\eeq
(The Wilson term mimicks a mass term although in a more subtle way; the mass
terms are "momentum dependent")
This leads to the modified propagator 
(in momentum space)
\beq
D_{w}(k)=\sum_{i=1}^{d}\gi (\frac{\sin k_{i}a}{a}) +m+\frac{r}{a}\sum_{i=1}^{d}
(1-\cos(k_{i}a))
\quad .
\eeq
In the above expression we have also introduced a bare mass term $m$.
The momentum dependent mass terms ensure that the modes at the corners of
the Brillouin zone
have masses of the order of $1/a$ and decouple from
the low energy effects  (in the limit $a\rightarrow 0$).
The price paid for eliminating these doublers
is the lack of chiral symmetry in the fermion action
(the Wilson term explicitly breaks chiral symmetry ).

In order to define the real time evolution of Ginsparg-Wilson fermions in the
Hamiltonian formulation, we have to first construct a Hamiltonian operator
starting from the Euclidean functional integral. In the transfer matrix
formalism this is done by choosing a particular axis as the time direction
(with a lattice spacing $\tau$)
and then taking the so called $\tau$ continuum limit. The $\tau$
continuum limit ($\tau \rightarrow 0$) 
is taken on an anisotropic lattice with different spacings in
the space and the time directions. In the Euclidean formulation, the Ginsparg-Wilson
operator can be formally understood to have arisen out of a chirally non-invariant blocking
transformation. One could have used a $\tau$-continuum procedure to obtain the
Hamiltonian version of the GW prescription. However, since we are eventually
interested only in the Hamiltonian, it is sufficient to
do the block spinning only in the spatial directions which already yields an
(anisotropic) lattice with a blocked action.
This blocked action has the same partition function as the
original action, though spatial correlation lengths are halved on this lattice.
We can then proceed to construct a Hamiltonian operator via the transfer
matrix in the usual way for fermions \cite{wilm}. This way we are able to
study the real time evolution of Ginsparg-Wilson fermions. 
As is well known, the
passage from the transfer-matrix formalism to the Hamiltonian formalism can
be carried out in more than one way, depending on the simplicity or
complexity one desires. 
Our Hamiltonian is invariant under the symmetry discussed
by Luscher in
\cite{luscher} and is a valid starting point
for carrying out calculations of wave functions, ground states, excited states,
etc.
The chirally invariant action on the lattice is 
\beq
S=\sum_{m,n}\bar \phi_{m} h_{mn}\phi_{n}
\quad ,
\eeq
where $h$ satisfies $\{h,\g5 \}=0$. We can block only the spatial degrees
of freedom by using the kernel (defined in \cite{gwils})
\beq
K(\ps,\phi)=\exp -\xi\sum_{m n}(\pb(m)-\bar \chi(m))(\ps(m)-\chi(m))
\quad .
\eeq
The only difference is that the blocked field $\chi(m)$ is defined as
\beq
\chi(m,t)=\eta \sum_{n}\phi (n,t) \quad ,
\eeq
where the above summation is only over fields defined over a hypercube at
the same instant. This is the condition which ensures that only the spatial
degrees of freedom are blocked. This leads to the following effective action
for the fermions on the blocked lattice
\beq
\exp (-\tilde A[\pb \ps])=\int d\bar \phi d\phi \exp [-A(\bar \phi, \phi)]
K(\ps,\phi)
\quad .
\eeq
The action on the blocked lattice is given by
\beq
\tilde A[\pb,\ps]=\sum_{m,n}\pb(m)\tilde h_{mn}^{\prime}\ps(n)
\quad ,
\eeq
where the fields $\ps$ are defined on the blocked lattice which has twice
the spacing (in the spatial direction) of the original lattice.
The propagator of the blocked lattice satisfies the following relations
\begin{eqnarray*}
\{\tilde h_{t},\g5 \}=0 \\ \nonumber
\{\tilde h_{s},\g5 \}= a \tilde h_{s}\g5 \tilde h_{s}
\quad .
\end{eqnarray*}
After the usual passage to the Hamiltonian via a transfer matrix formalism
the Hamiltonian one obtains is simply
\beq
H=\sum_{x,y} \psi^{\dagger}(x) \gamma_{0}\tilde h_{s}(x,y)\psi(y)
\quad .
\eeq

Henceforth,
the operator $\tilde h_{s}$ will be called $D$ and it satisfies the G-W
relation. Any $\tilde h_{s}$ satisfying the Ginsparg-Wilson relation can be used
for defining our Hamiltonian.
An explicit choice for $D$ satisfying the Ginsparg-Wilson relation is
the overlap operator
given by Neuberger`s construction \cite{neub}
\beq
D=\frac{1}{a} ( 1-\frac{A}{\sqrt{A^{\dagger} A}} )
\quad ;
\eeq
A is defined in terms of the Wilson operator $D_{w}$ as
\beq
A=1-aD_{w}
\quad .
\label{wilmass}
\eeq
It can be easily checked that the operator $D$ satisfies the
G-W relation.
It is worth adding here that any $A$ satisfying the following properties:
$A^{\dagger}=\gamma_5 A \gamma_5$ and $A^{\dagger}A$ commutes with $A$ will
give a $D$ satisfying the GW relation. This may be useful in a more general
context. 

Though any $D$ satisfying the Ginsparg-Wilson relation can be used
to construct our Hamiltonian, we have used the above explicit form 
proposed by Neuberger et al for our
calculations. It should be stressed that apart from this choice, the
considerations of this paper are independent of the overlap formalism.
The above operator relations can be translated into momentum
space and it is in momentum space that we will make most of our
manipulations. 
It should be emphasised that in general, with external fields, momentum
space description is not very economical.But in the $1+1$ dimensional abelian
case with only an electric field, and in the $3+1$ dimensional case with 
uniform electric and magnetic fields, momentum space description is 
still useful.

The Hamiltonian of the lattice field theory is
\beq
H=\sum_{x,y} \pb(x)D(x,y)\ps(y)
\quad .
\eeq
$\pb$ and $\ps$ can be interpreted as field operators in the usual sense.
The wave equation for the fermion fields is
\beq
i\frac{\partial \ps(x,t)}{\partial t}=\sum_{y}\gamma_{0} D(x,y)\ps(y)
\quad .
\label{deqn}
\eeq
Using the properties of $D$ and $\gamma_{0}$ it is easy to show that the
Hamiltonian is hermitian, and therefore the evolution is unitary.
We are basically interested in how the anomaly arises in this model.
To get the anomaly we must of course couple the
fermions to an external gauge field and then look for non-conservation
of the chiral charge. 
It is well known that the anomaly can be extracted
by treating the gauge fields as a classical variable and quantizing only
the fermions. In order to be able to study the problem of fermions in an
external field we will have to make some approximations which will be
described shortly. The analysis will be presented in 1+1 dimensions.

In 1+1 dimensions the only effect of the external gauge field is to shift
the momentum variable $k$ to $k-ga(t)$ where $a(t)$ is the time dependent
component of the vector potential ($A_{1}(x,t)=a(t)\ and A_{0}=0$).
This is true only in the Hamiltonian picture, even in the $1+1$ case,
because in the Euclidean case, $D^{\dagger}D = (k-ea(t))^2 + \gamma_5 {\cal E}$,
where ${\cal E}$ is the electric field.
We define the chiral charge operator
on the lattice in the usual
manner as
\beq
Q_{n}=a\ \sum_{x}\ps^{\dagger}(x,t) \g5 \ps(x,t)
\quad .
\eeq
All the operators are defined in the Heisenberg representation.
The time dependence of the chiral charge is given by
\beq
\dot Q_{n}=\frac{\partial Q_{n}}{\partial t}+i[H,Q_{n}]
\quad .
\eeq
Since only the second term on the righthand side contributes to
$\dot Q_{n}$ we have
\beq
\dot Q_{n}=i[H,Q_{n}] \quad .
\eeq
Evaluating the commutator this becomes
\beq
\dot Q_{n}=i\sum_{x,y}\pb (x,t) \{ D,\g5 \} \ps(x,t)
\quad .
\eeq
Using the Ginsparg-Wilson relation this becomes
\beq
\dot Q_{n}=i a\sum_{x,y}\pb(x,t) D \g5 D \ps(y,t)
\quad .
\eeq
The spatial indices of $D$ and all Dirac indices
have been suppressed for ease in reading.
Since $D\g5 D$ is non-zero we see that the chiral charge is not in general
conserved, as expected. It remains to be seen if this non-conservation
of the chiral charge reproduces the correct anomaly.
For this the Dirac equation in Eq.~\ref{deqn} 
has to be solved in the presence of an
external potential and the solutions have to be examined.
We consider a time dependent potential which rises from zero to a constant
value $A_{\tau}$ in a time $\tau$. $\tau$ is a time scale in the problem
and two cases can be easily analyzed, 
the sudden limit $\tau \rightarrow 0$ and
the adiabatic limit $\tau \rightarrow \infty$, as was also done in \cite{amb}.
First one writes the free field $\ps(x,t)$ as a superposition of positive and
negative energy spinors
\beq
\ps(x,t)=\int_{BZ} \frac{1}{2\pi} [b(k)u(k)\exp (-ikx)+d^{\dagger}(k)v(k)\exp (ikx)]
\quad .
\eeq
$kx$ is short for $k_{0}x_{0}-k_{1}x_{1}$. 
$E=k_{0}$. As usual $u(k)$ and $v(k)$
represent positive and negative energy spinors. The operator $b(k)$ destroys
an electron of momentum $k_{1}$ and the operator $d^{\dagger}(k)$ creates a
positron of momentum $-k_{1}$. 
Putting these spinors in the equation for the axial charge we have
\beq
\langle |\dot Q_{n}|\rangle=\int_{bz} dk \frac{1}{2\pi}\bar v(k,t) D(k,t)\g5 
D(k,t) v(k,t)
\langle d(k) d^{\dagger}(k) \rangle
\quad .
\label{anom}
\eeq
The angular brackets denote expectation values in the vacuum which is
the state with zero electrons and positrons.
Using the definition of $D$ and the properties of the gamma matrices it is
easy to show that 
\beq
D(k)\g5 D(k)=\g5 D^{\dagger}(k) D(k)
\quad .
\eeq
Now , $D^{\dagger}(k) D(k)$ is a c-number and acts trivially on the
Dirac spinors.
($D^{\dagger}D$ is not a c-number in general.For the specific class where
$D$ is of the form $A_i\gamma_i+B$ with $A_i,B$ commuting, this is true.Already
in the $3+1$ dimensional case this is no longer true even for a uniform magnetic 
field.)
In order to evaluate $\bar v(k,t) \g5 v(k,t)$ we have to determine the
evolution of the negative energy spinor in the external field $a(t)$.
Before we calculate this quantity it is instructive to calculate the same
without any field. In the absence of an external field the positive and negative energy states evolve as 
\begin{eqnarray*}
u(k,t)=\exp (-iE(k)t) \tilde u(k) \\ \nonumber
v(k,t)=\exp (iE(k)t) \tilde v(k)
\quad .
\end{eqnarray*}
The spinors $\tilde u(k)$ and $\tilde v(k)$ satisfy the time independent
Schroedinger equations with positive and negative energies.
\begin{eqnarray*}
\gamma_{0}D(k) \tilde u(k)=E(k) \tilde u(k) \\ \nonumber
\gamma_{0}D(-k) \tilde v(k)=-E(k) \tilde v(k)
\quad .
\end{eqnarray*}
The Weyl representation for the $2$ dimensional $\gamma$ matrices 
is
\beq
\gamma_{1}=i \gamma_{5}=i \sigma_{3} \quad \quad
\gamma_{0}=\sigma_{1}
\quad \gamma_{5}=\sigma_{2}
\quad .
\eeq
$\sigma_{1}$ and $\sigma_{3}$ are the Pauli matrices.
The eigen values of the spinors are given by
$E(k)^2 =D^{\dagger}(k)D(k)$.
Using the definition of $D(k)$ we get $D^{\dagger}(k)D(k)$ to be a c-number.
$D(k)$ can be written as
\beq
D(k)=\frac{1}{a} (1-\frac{(1-am(k))}{\sqrt(A^{2}(k))})
+  \frac{1}{a}(\frac{\sin (k_{1}a)}{\sqrt(A^{2}(k))}) \gamma_{1}
\quad .
\eeq
If we now write $D(k)$ as
\beq
D(k)=g(k,t)+\gamma_{1}f(k,t)
\quad ,
\eeq
then
$E^{2}(k)=D^{\dagger}(k)D(k)=f^{2}(k,t)+g^{2}(k,t)$.
$f$ and $g$ are functions given by
\begin{eqnarray*}
g(k)=\frac{1}{a} (1-\frac{(1-am(k))}{\sqrt(A^{2}(k))}) \\ \nonumber
f(k)=\frac{1}{a}(\frac{\sin (k_{1}a)}{\sqrt(A^{2}(k))}) \\ \nonumber
\quad .
\end{eqnarray*}
The function $f$ is an odd function of $k$ whereas the function $g$ is
an even function of $k$.
The time independent spinors can be normalized to satisfy
\begin{eqnarray*}
\bar {\tilde u(k)} \g5 \tilde u(k)=0 \\ \nonumber
\bar {\tilde v(k)} \g5 \tilde v(k)=0 \\ \nonumber
\bar {\tilde u(k)} \g5 \tilde v(k)=1
\quad .
\end{eqnarray*}
This means that $\langle \dot Q_{n} \rangle=0$ in the absence of an
external field.
This is as expected, there is no anomaly in zero 
external field even though the Hamiltonian is not chirally invariant.

In the presence of an external field the only
change in the operator $D$ is a replacement of $k$ by $k-g\ a(t)$.
Since the structure of $D$ is not affected by an external field the
Ginsparg-Wilson relation is still satisfied. Although the evolution of
Dirac spinors in an arbitrary external field can only be analyzed
numerically, two limiting cases admit a simpler analysis. These
are the adiabatic limit and the sudden limit, and we shall examine these
two cases separately.
When an external field is turned on slowly  (compared to the time scales in
the system) we can use the adiabatic approximation. The sudden approximation
is useful when the field is turned on faster than the fastest time scale in the system.
The rate at which the field is turned on can be controlled by introducing
a parameter $\tau$ defined as follows
\beq
A(t)=0 \quad t\le 0 \nonumber
\eeq
\beq
A(t)=A \quad t > \tau
\quad .
\eeq
The precise form of $A(t)$ for $0<t<\tau$ is not very important.
The sudden limit corresponds to $\tau \rightarrow 0$ and the adiabatic
limit corresponds to $\tau \rightarrow \infty$.
In the adiabatic approximation the
form of the positive and negative energy
spinors for times $0<t<\tau$ is given by
\beq
u(k,t)=a(k,t)\xi^{\star}(t)u^{0}(k,t)+b(k,t)\xi(t)v^{0}(k,t)
\label{adia1}
\eeq
\beq
v(k,t)=c(k,t)\xi^{\star}(t)u^{0}(k,t)+d(k,t)\xi(t)v^{0}(k,t)
\quad .
\label{adia2}
\eeq
The above equation is written for the individual fourier components of the
positive and negative energy spinors in an external field,
$u^{0}(k,t)$
 and $v^{0}(k,t)$ 
are the positive and negative energy spinors satisfied by the
instanteneous Schrodinger equation at the instant $t$.
We closely follow the notation of \cite{amb} and we have also corrected some of
the misprints which occur therein.
$\xi(t)^{\star}$ and $\xi(t)$ are the phase factors for the positive
and negative energy spinors.
The
phase factor $\xi(t)$ is given by
\beq
\xi(t)=\exp (i\int_{0}^{t}E(k,t^{\prime})d t^{\prime}
\quad .
\eeq
$a,b,c,\  and\  d$ are called Bogoulobov coefficients. 
The Bogoulobov co-efficients are chosen to satisfy 
the following boundary conditions
\beq
a(0)=d(0)=1 \quad \quad b(0)=c(0)=0 \quad .
\eeq
These boundary conditions ensure that we are looking at the evolution
of the positive and the negative energy states before the external field
is switched on.
As mentioned before, $u^{0}(k,t)$ and $v^{0}(k,t)$ are positive
and negative energy spinors having momentum $k$ and $-k$ respectively, and they
satisfy the instanteneous Schrodinger equations given by
\begin{eqnarray*}
\gamma_{0}D(k) u^{0}(k,t)=E(k,t) u^{0}(k,t) \\ \nonumber
\gamma_{0}D(-k) v^{0}(k,t)=-E(k,t) v^{0}(k,t)
\quad .
\end{eqnarray*}
Substituting the expressions in Eq.~\ref{adia1} and Eq.~\ref{adia2} in the wave equation for the
fermions
and using the previously mentioned boundary conditions we get
\beq
c(k,t)=\int_{0}^{t}\alpha(k,t^{\prime})d(k,t^{\prime})\xi^{2}(t^{\prime})d
t^{\prime}
\label{a1}
\eeq
\beq
d(k,t)=1-\int_{0}^{t}\alpha(k,t^{\prime})c(k,t^{\prime}){\xi^{\star}}^{2}
(t^{\prime})dt^{\prime}
\label{a2}
\eeq
\beq
b(k,t)=-c^{\star}(k,t)
\label{a3}
\eeq
\beq
a(k,t)=d^{\star}(k,t)
\quad .
\label{a4}
\eeq
The quantity
$\alpha(k,t)$ is defined by
\beq
\dot u^{0}(k,t)=\alpha(k,t) v^{0}(k,t)
\quad ,
\eeq
and is
\beq
\alpha(k,t)=\frac{(g \dot f-f \dot g)}{2 E^{2}(k,t)}
\quad .
\eeq
After using the stated normalizations of the spinors, 
$\bar v(k,t)\g5 v(k,t)$ is given
\beq
\bar v(k,t) \g5 v(k,t)=c^{\star}d\xi^{2} -c.c
\quad .
\eeq
The co-efficients $c(k,t)$ and $d(k,t)$ can be approximated by (for small
values of $\alpha(k,t)$)
\beq
c(k,t)=-i\frac{\alpha(k,t)}{2 E(k,t)}\xi^{2}(t) \quad
d(k,t)=1+ O(\alpha^2)
\quad .
\eeq
In the adiabatic approximation the quantities inside the integrand
on the right hand side of
Eq.~\ref{a1} and Eq.~\ref{a2} are
evaluated at $t=0$.
Substituting for the values of $f(k,t)$ and $g(k,t)$ and using the 
relation
\beq
D(k) \g5 D(k)=\g5 D^{\dagger}(k) D(k)=\g5 E^{2}(k)
\eeq
the r.h.s of Eq.~\ref{anom} becomes
\beq
\int_{BZ}\frac{1}{2\pi}  dk\  C(k)\  (a\  a\  g \dot a(t))
\eeq
where $C(k)$, a complicated expression, is given in the appendix
along with the expressions for $f,g,\dot f,\dot g$. The function $C(k)$
can be plotted and the integral of $C(k)$ over the Brillouin zone can be
estimated numerically.
The function $C(k)$ depends on $r,k,a,m$. We plot $C(k)$ as a function of
$k$ in Fig.~\ref{r0} to Fig.~\ref{r5.0}. We first plot it for zero mass
and then for a non-zero value ($m=5$).
The eigenvalues of the spinors (in a zero external field) are
also plotted in Fig.~\ref{e0.0} to Fig.~\ref{e5.0}.
The first thing we observe is that when $r=0$ the eigenvalue spectrum does
not distinguish between the modes at $k=0$ and $k=\pi/a$, and the integral
of $C(k)$ over
the Brillouin zone is zero. For $r\ne 0$ the modes at $k=0$ and $k=\pi/a$ have
different energies and an asymmetry develops in the function $C(k)$.
For $r=1$, the integral of $C(k)$ over the Brillouin Zone
has the value $-2$.
Substituting this in Eq.~\ref{anom}
we get the anomaly equation
\beq
\langle \dot Q_{n} \rangle =-\frac{g}{2\pi} \int d^{2}x \epsilon_{\mu \nu} F_{\mu \nu}
\eeq
in the continuum limit.

We have studied the function $C(k)$ for different values of $r$ and we find
that
the anomaly is independent
of $r$ for large $r$ but vanishes for a smaller and, in particular, a zero value
of $r$. The value $r=0.5$ seems to separate the region with and without
the anomaly.
A non-zero Wilson $r$ parameter is necessary to
get the anomaly in the continuum limit. This means that a Neuberger like
operator for $D$ where the naive Dirac operator is used in 
place of $D_{w}$ in Eq.~\ref{wilmass} will not reproduce the correct
anomaly in the continuum limit inspite of satisfying the Ginsparg-Wilson
relation (the operator $D$ with $D_{w}$ replaced by $D_{naive}$ also
satisfies the Ginsparg-Wilson relation). The case $m\ne 0$ can also be
analyzed along the same lines and it turns out that the integral of
$C(k)$ is very small, consistent with zero. It appears that in this
case we have an exact cancellation of the bare mass term with the
anomaly term to give a zero rate of change of chiral charge. Nevertheless, the
anomaly term is still present and so is the mass term, but the two
appear with opposite signs.
The adiabatic approximation is justified when the switching time $\tau$
is much greater than the characteristic time periods of the system.
In our example $2\pi/E(k)$ is the characteristic time period of the system
and the adiabatic approximation is justified when $\tau >> a$. To summarize,
a zero bare mass term with a non-zero $r$ parameter gives an anomaly
independent of $r$ for $r>0.5$.
When a bare mass term is included we have a cancellation of
the anomaly term with the mass term though both terms are still present.

The rate of change of the chiral charge can also be calculated in the
sudden approximation in which an external field is turned on infinitely fast.
This approximation corresponds to the limit $\tau \rightarrow 0$. In this
limit the spinors $u^{0}(k,t),v^{0}(k,t)$ are unchanged immediately after the field is
turned on and only their evolution is governed by the new
Hamiltonian (with a constant field). The normalization of the spinors
in a constant external field (with value $A_{\tau}$) 
can be made just as in the zero field case and
there is no anomaly in this limit.
This limit corresponds to the case $\tau <<a$.

The time $\tau$ is a characteristic time associated with the gauge fields
on the lattice and our calculation clearly shows that in order to get the
correct anomaly we have to ensure that we are not in the regime of the
sudden approximation. If we are in the intermediate region we will see
a crossover from one limit to another limit.
So far we have only studied fermions and (abelian) gauge fields
in 1+1 dimensions. It was pointed out in \cite{niels} that the anomaly in
3+1 dimensions factorizes into a 1+1 dimensional part and an extra factor
coming from the additional dimensions.
We briefly review the argument
in \cite{niels}. To get the anomaly in 3+1 dimensions we first turn on a 
magnetic field in the, say z, direction. This leads to the usual Landau
levels for the fermions which are labelled by integers. We then turn on
an electric field paralell to the magnetic field. The important point is that
the fermions in the lowest Landau level in the presence of this
electric field behave like fermions in the 1+1 dimensional case that we
have just analyzed. Hence the same 1+1 dimensional anomaly is present but
with an additional degeneracy factor coming from the Landau levels. 
As shown in \cite{amb} the argument goes through for the lattice Dirac sea
case for the case of an uniform magnetic field.
The degeneracy factor is a 
geometrical
quantity that is in general dependent on the details of the lattice Hamiltonian which
is more complicated for Ginsparg-Wilson fermions. 
However, when the
Ginsparg-Wilson operator is constructed as $aD_{GW}= (1- {A\over{\sqrt{A^{\dagger}A}}})$
with $A = 1- aD_W$, the degeneracies of $D_{GW}$ and $D_W$ are the same, and in the limit
of zero lattice spacing
the degeneracy is just 
\beq
L_{1}L_{2}g H/(2\pi)
\quad ,
\eeq
the number of states in the square $L_{1}L_{2}$ perpendicular to the
magnetic field ($H$). The above factor simply multiplies the $1+1$ dimensional
anomaly and gives the correct anomaly in 3+1 dimensions.

The main aim of this note was to show that a Hamiltonian analysis of
Ginsparg-Wilson fermions leads to a non-zero rate of chiral charge and
gives the anomaly equation in the continuum limit. The doubler modes
are suppressed by the Wilson parameter, infact the Wilson
parameter $r$ plays a crucial role in yielding the correct anomaly.
A quantum mechanical analysis supplemented by an adiabatic approximation
was necessary to get the anomaly. It is noteworthy that if we are not in
the adiabatic regime we will get other contributions ($\dot a(t)$ and higher
time derivatives) to the chiral charge and this will not reproduce
the anomaly equation. It may be useful to compare our derivation with that of
the overlap method. In the overlap method the anomaly is extracted by looking
at the scalar product of the ground states of two different many 
body Hamiltonians, whereas in our approach we study the dynamical picture behind
the anomaly by using the properties of the
Dirac sea in an external electric field in the adiabatic limit.
\newpage
Appendix

In this appendix we collect together some expressions which are necessary
to get the function $C(k)$.
\beq
f(k)={\frac{1 - {\frac{1 - a\,\left( m + 
            {\frac{r\,\left( 1 - \cos (a\,k) \right) }{a}} \right) }{{
           \sqrt{1 - 2\,a\,\left( m + 
               {\frac{r\,\left( 1 - \cos (a\,k) \right) }{a}} \right)  + 
            {a^2}\,{{\left( m + 
                  {\frac{r\,\left( 1 - \cos (a\,k) \right) }{a}} \right) }^2}
             + {{\sin (a\,k)}^2}}}}}}{a}}
\eeq
\beq
g(k)=
{\frac{\sin (a\,k)}
   {{a}\,{\sqrt{1 - 2\,a\,
          \left( m + {\frac{r\,\left( 1 - \cos (a\,k) \right) }{a}} \right)  +
          {a^2}\,{{\left( m + {\frac{r\,\left( 1 - \cos (a\,k) \right) }
                 {a}} \right) }^2} + {{\sin (a\,k)}^2}}}}}
\eeq
\begin{eqnarray*}
\dot f(k)=
{\frac{\left( 1 - m - {\frac{r\,\left( 1 - \cos (a\,k) \right) }{a}} \right)
        \,\left( -2\,r\,\sin (a\,k) + 
        2\,a\,r\,\left( m + {\frac{r\,\left( 1 - \cos (a\,k) \right) }
             {a}} \right) \,\sin (a\,k) + 2\,\cos (a\,k)\,\sin (a\,k) \right) 
      }{2\,a\,{{\left( 1 - 2\,a\,
            \left( m + {\frac{r\,\left( 1 - \cos (a\,k) \right) }{a}} \right) 
            + {a^2}\,{{\left( m + 
                 {\frac{r\,\left( 1 - \cos (a\,k) \right) }{a}} \right) }^2} +
            {{\sin (a\,k)}^2} \right) }^{{\frac{3}{2}}}}}} + \\
  {\frac{r\,\sin (a\,k)}
    {a\,\left( 1 - 2\,a\,\left( m + 
           {\frac{r\,\left( 1 - \cos (a\,k) \right) }{a}} \right)  + 
        {a^2}\,{{\left( m + {\frac{r\,\left( 1 - \cos (a\,k) \right) }
                {a}} \right) }^2} + {{\sin (a\,k)}^2} \right) }}
\end{eqnarray*}
\begin{eqnarray*}
\dot g(k)=
{\frac{-\sin (a\,k)}
    {2\,a\,{{\left( 1 - 2\,a\,\left( m + 
              {\frac{r\,\left( 1 - \cos (a\,k) \right) }{a}} \right)  + 
           {a^2}\,\left( m + {\frac{r\,\left( 1 - \cos (a\,k) \right) }
                {a}} \right)  + {{\sin (a\,k)}^2} \right) }^{{\frac{3}{2}}}}}}
\\
   + {\frac{\cos (a\,k)}
    {{a}\,\left( 1 - 2\,a\,
         \left( m + {\frac{r\,\left( 1 - \cos (a\,k) \right) }{a}} \right)  + 
        {a^2}\,\left( m + {\frac{r\,\left( 1 - \cos (a\,k) \right) }{a}}
            \right)  + {{\sin (a\,k)}^2} \right) }}
\end{eqnarray*}
\begin{eqnarray*}
E^{2}(k)=
{\frac{{{\sin (a\,k)}^2}}
    {{a^2}\,\left( 1 - 2\,a\,\left( m + 
           {\frac{r\,\left( 1 - \cos (a\,k) \right) }{a}} \right)  + 
        {a^2}\,{{\left( m + {\frac{r\,\left( 1 - \cos (a\,k) \right) }
                {a}} \right) }^2} + {{\sin (a\,k)}^2} \right) }} + 
\\
  {\frac{{{\left( 1 - {\frac{1 - 
              a\,\left( m + {\frac{r\,\left( 1 - \cos (a\,k) \right) }
                   {a}} \right) }{{\sqrt{1 - 
                 2\,a\,\left( m + 
                    {\frac{r\,\left( 1 - \cos (a\,k) \right) }{a}} \right)  + 
                 {a^2}\,{{\left( m + 
                       {\frac{r\,\left( 1 - \cos (a\,k) \right) }{a}} \right) 
                      }^2} + {{\sin (a\,k)}^2}}}}} \right) }^2}}{{a^2}}}
\end{eqnarray*}
\beq
C(k)= (1/(2 E(k)) (g(k) \dot f(k) -f(k) \dot g(k))
\eeq

\newpage
\begin{figure}
\label{r0}
\caption{$C(k)$ at m=0 and r=0.}
\end{figure}
\begin{figure}
\label{r0.4}
\caption{$C(k)$ at m=0 and r=0.4}
\end{figure}
\begin{figure}
\label{r0.6}
\caption{$C(k)$ at m=0 and r=0.6}
\end{figure}
\begin{figure}
\label{r1.0}
\caption{$C(k)$ at m=0 and r=1.0}
\end{figure}
\begin{figure}
\label{r5.0}
\caption{$C(k)$ at m=0 and r=5.0}
\end{figure}
\begin{figure}
\label{e0.0}
\caption{$E(k)$ at m=0 and r=0.0}
\end{figure}
\begin{figure}
\label{e0.4}
\caption{$E(k)$ at m=0 and r=0.4}
\end{figure}
\begin{figure}
\label{e0.6}
\caption{$E(k)$ at m=0 and r=0.6}
\end{figure}
\begin{figure}
\label{e1.0}
\caption{$E(k)$ at m=0 and r=1.0}
\end{figure}
\begin{figure}
\label{e5.0}
\caption{$E(k)$ at m=0 and r=5.0}
\end{figure}
\begin{figure}
\label{m5r0}
\caption{$E(k)$ at m=5 r=0}
\end{figure}
\begin{figure}
\label{m5r5}
\caption{$E(k)$ at m=5 r=5}
\end{figure}
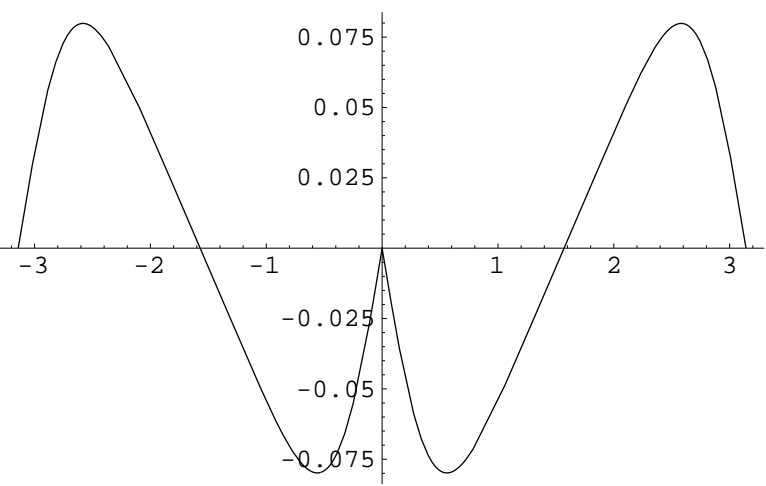
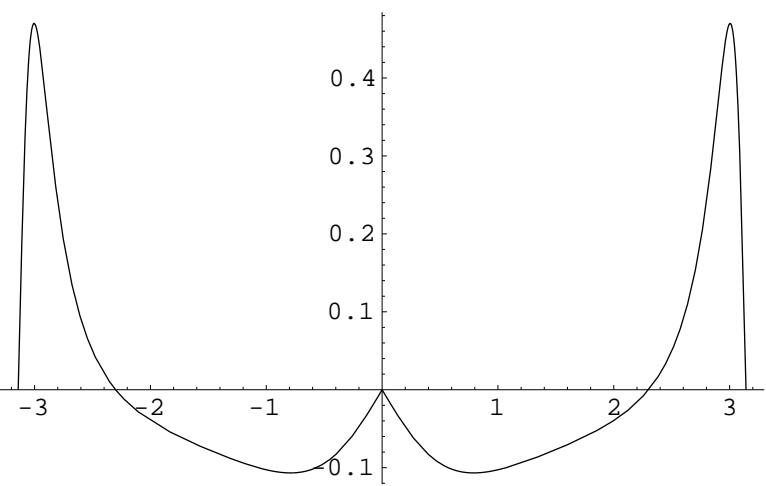
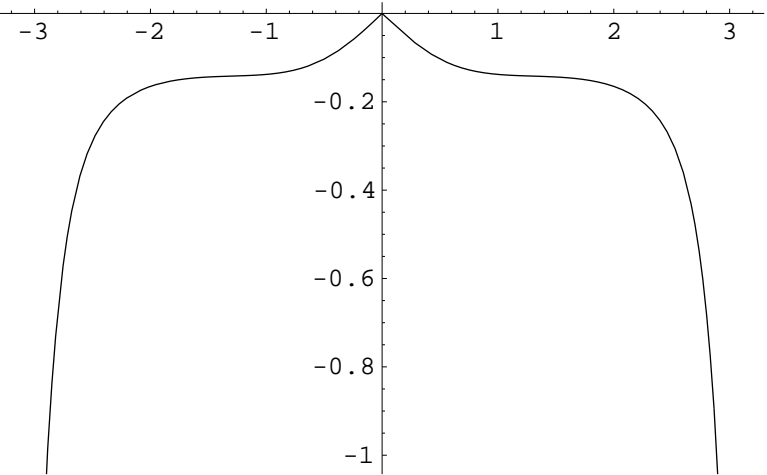
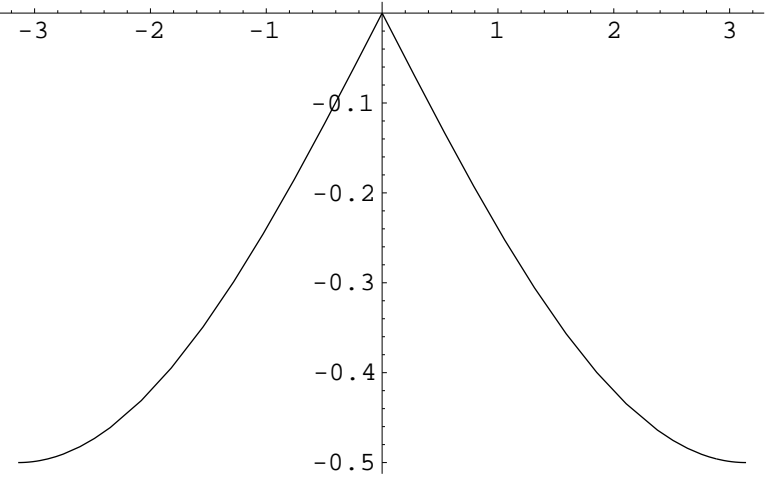
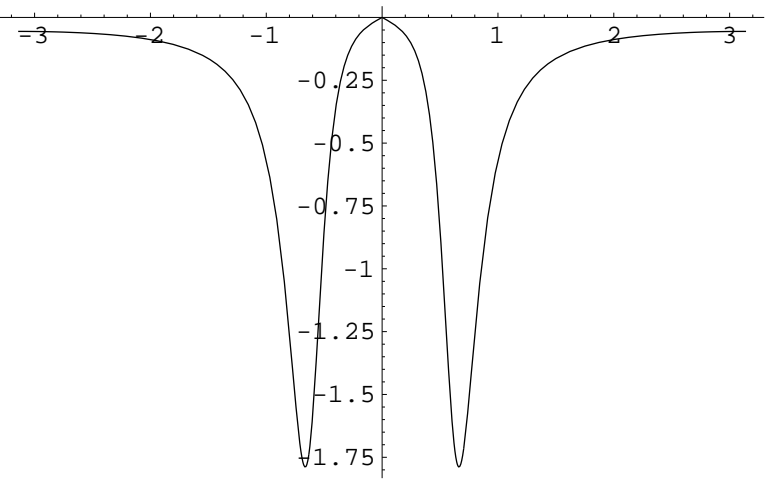
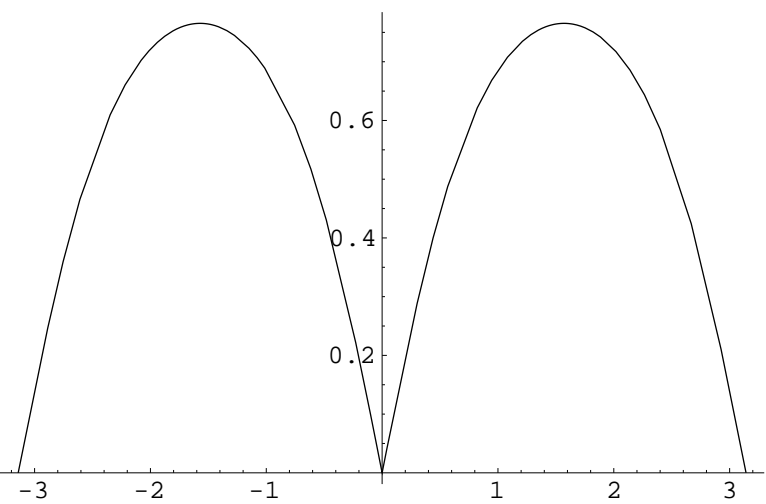
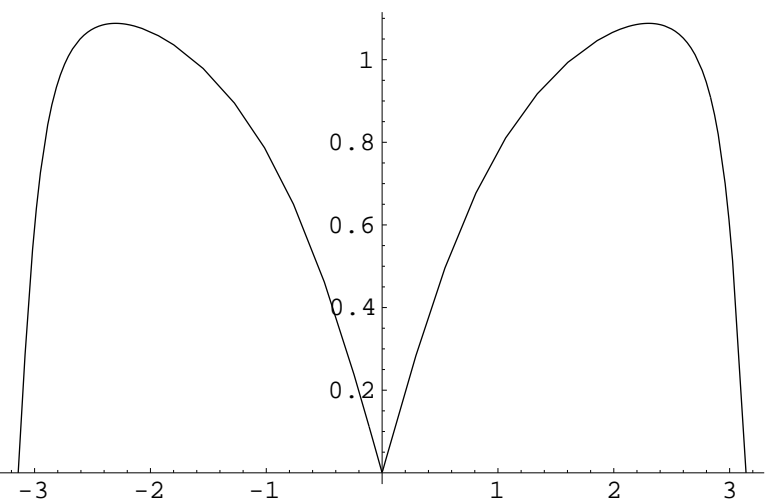
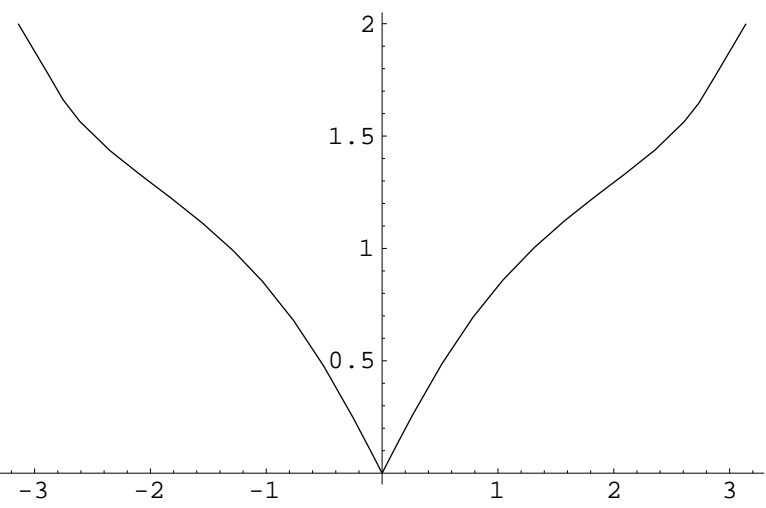
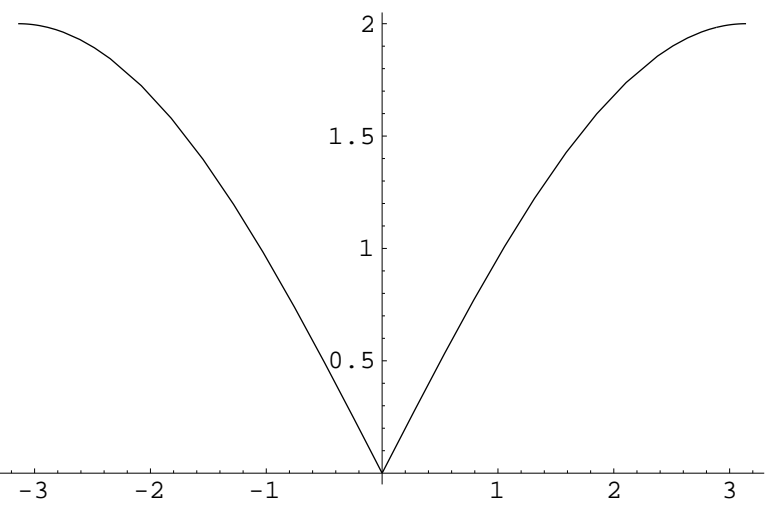
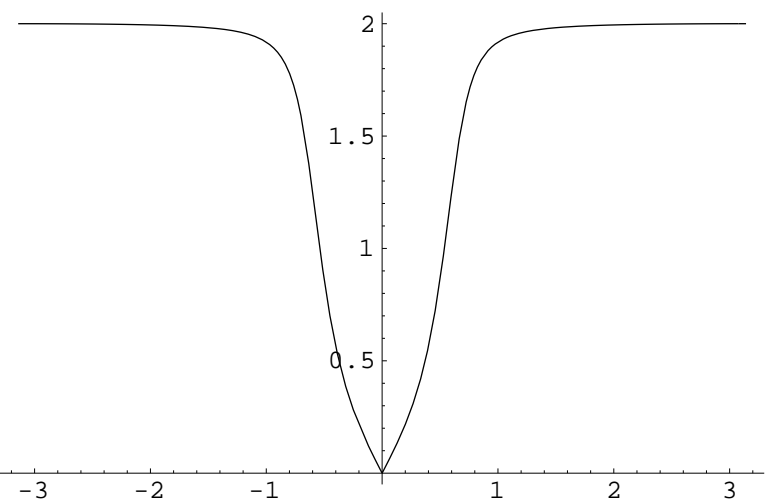
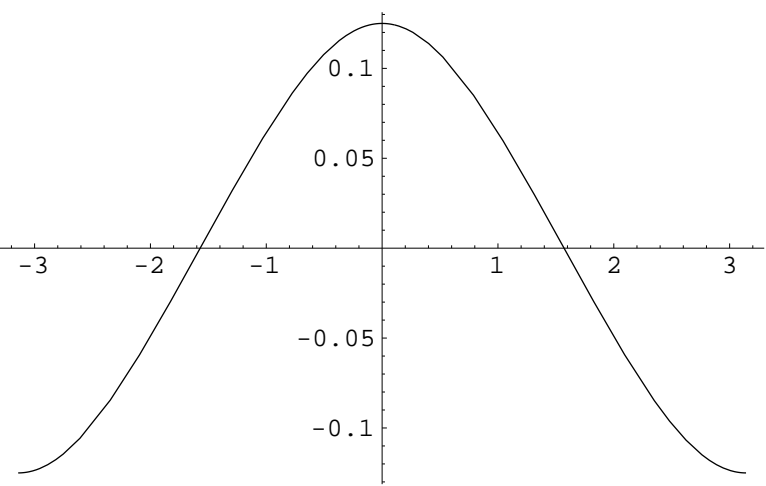
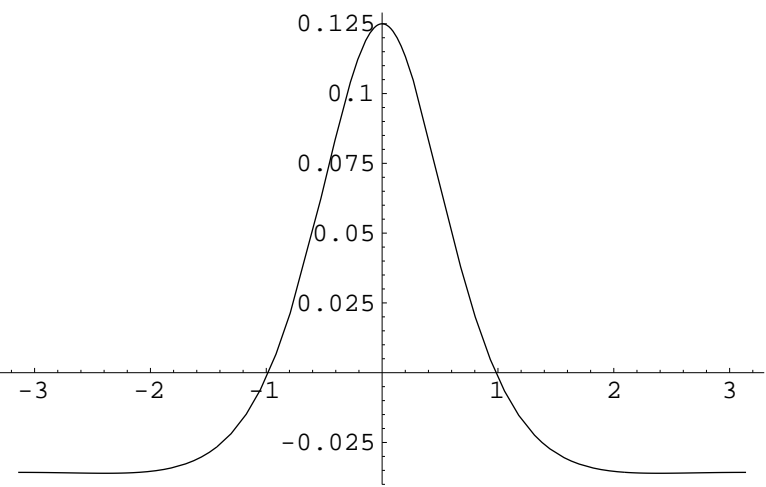
\end{document}